\documentclass[12pt,superscriptaddress,aps,prd,preprint,showpacs]{revtex4}

\usepackage[dvips]{graphicx}
\usepackage{amsfonts}
\usepackage{slashed}
\usepackage[utf8]{inputenc}
\usepackage{amsmath}
\usepackage{hyperref}
\usepackage{xcolor}
\usepackage{mathtools}

\begin{document}

\title{Dynamical symmetry breaking in a non-polynomial four-fermion model}

\author{J. F. Assun\c c\~ao}
\affiliation{Departamento de F\'\i sica, Universidade Estadual de Alagoas,\\ 57025-602, Macei\'o, Alagoas, Brazil}
\email{jose.assuncao@uneal.edu.br}
\author{T. Mariz}
\affiliation{Instituto de F\'\i sica, Universidade Federal de Alagoas,\\ 57072-900, Macei\'o, Alagoas, Brazil}
\email{victor.prata,tmariz@fis.ufal.br}
\author{A. Yu. Petrov}
\affiliation{Departamento de F\'\i sica, Universidade Federal da Para\'{\i}ba,\\
 Caixa Postal 5008, 58051-970, Jo\~ao Pessoa, Para\'\i ba, Brazil}
\email{petrov@fisica.ufpb.br}
\author{V. Prata}
\affiliation{Instituto de F\'\i sica, Universidade Federal de Alagoas,\\ 57072-900, Macei\'o, Alagoas, Brazil}
\email{victor.prata,tmariz@fis.ufal.br}

\begin{abstract}

We present a new four-fermion model with a strongly non-polynomial coupling. To eliminate these non-polynomialities, we introduce two auxiliary vector fields and show that the dynamical Lorentz symmetry breaking is possible in our theory, as revealed by the effective potential generated for these vector fields. We explicitly calculate the low-energy effective action for these fields, which is given by the sum of three Maxwell terms (one for each field and a mixed one) plus a potential term displaying nontrivial minima.

\end{abstract}


\maketitle

\section{Introduction}

The idea of dynamical symmetry breaking in a four-fermion model was first introduced by Bjorken in Ref.~\cite{Bjorken:1963vg} (see also the more recent review in Ref.~\cite{Bjorken:2001pe}). Later, Amati et al.~\cite{Amati:1981xt} proposed a non-polynomial version of this four-fermion model, and Garavaglia subsequently investigated the vacuum structure of the vector fields~\cite{Garavaglia:1986cy}. One of the main motivations for this concept is the idea that various field theory models, in particular gauge ones, can be treated as emergent phenomena, thus providing a mechanism to explain the origin of the observed physical interactions. Besides the vector field theories studied in Refs.~\cite{Bjorken:1963vg,Amati:1981xt,Garavaglia:1986cy}, this also includes gravity~\cite{Verlinde:2016toy}. It is worth mentioning that gauge field theories were also shown to arise dynamically in $CP^{n-1}$ models~\cite{DAdda:1978dle}.

The special interest in the dynamical generation of symmetry-breaking theories is related to the use of the bumblebee model, originally proposed in \cite{Kostelecky:2000mm,Kostelecky:2003fs}, where the Lorentz symmetry breaking (LSB) occurs spontaneously from a potential displaying an infinite set of minima. It is worth mentioning that the mechanism of spontaneous LSB is the most appropriate one for the case of a curved space-time \cite{Kostelecky:2003fs}, since, in addition to the natural origin of the LV vector (in general, tensor), this mechanism allows one to avoid serious difficulties related to implementing explicit LSB in a curved space-time, such as, for example, the necessity of no-go constraints for LV vectors (tensors) (see the discussion in \cite{Kostelecky:2020hbb}). Various aspects of the bumblebee theory in a curved space-time, that is, of bumblebee gravity, have been studied earlier; for example, some exact gravitational solutions were found, see e.g. \cite{Bertolami:2005bh,Casana:2017jkc,Ovgun:2018xys,Maluf:2020kgf,Belchior:2025xam} and many other papers. 

At the same time, studies of the bumblebee theory in a flat spacetime are still very interesting and important. Examples of interesting studies along this direction are presented in \cite{Maluf:2014dpa}, where the dispersion relation of the graviton on a flat background in the presence of the bumblebee field is found, and in \cite{Hernaski:2014jsa,Maluf:2015hda,Escobar:2017fdi}, where issues related to quantization of the bumblebee fields are discussed. One of the interesting effects related to this theory is the possibility of its dynamical generation, i.e., the bumblebee theory with such a potential can arise dynamically as a one-loop correction in a special four-fermion model, as was proved in \cite{Gomes:2007mq}. Further, another version of the four-fermion model allowing for the generation of the bumblebee potential, describing massless fields, was presented in \cite{Assuncao:2017tnz} (the method of calculations presented in that paper was generalized for the massive case in \cite{Araujo:2024tiy}), and in \cite{Assuncao:2019azw}, this methodology was applied to the tensor bumblebee model.

In this paper, we take the next step: using the approach presented in \cite{Garavaglia:1986cy}, based on non-polynomial couplings of fermions, we introduce not one, but two vector fields and study their emergent dynamics, with the potential turning out to be bounded from below and to display an infinite set of minima, just as the bumblebee model requires.

The paper is organized as follows. In Section \ref{potential}, we calculate the effective potential of our theory and demonstrate that it is bounded from below and displays an infinite set of minima. In Section \ref{action}, we calculate the kinetic terms of our effective action. Finally, in Section \ref{summary}, we discuss the results and perspectives.

\section{non-polynomial four-fermion model and emergent vector fields}\label{potential}

The four-fermion model of interest is described by the Lagrangian similar to the one introduced in \cite{Garavaglia:1986cy}:
\begin{equation}\label{L0}
\mathcal{L}_0 = \bar{\psi}i\slashed{\partial}\psi - eg(\bar{\psi}\gamma^\mu \psi) (\bar{\psi}i\overset{\leftrightarrow}{\partial}_{\!\mu}\psi /\bar{\psi}\psi),
\end{equation}
where $e$ and $g$ are coupling constants, and $\overset{\leftrightarrow}{\partial}_{\!\mu} = \partial_\mu - \overset{\leftarrow}{\partial}_{\!\mu}$ is the left-right derivative. In principle, this Lagrangian is locally gauge invariant under the transformation $\psi \to e^{i\alpha(x)} \psi$, provided that the condition $1-2eg=0$ is satisfied. However, this is a nonperturbative condition since the smallness of $e$ requires a large value of $g$ and vice versa. To avoid potential difficulties with the use of the perturbative expansion, we will further assume that our fields are small, which is natural within the perturbative framework, so that at one-loop order the smallness of quantum corrections is enforced by the smallness of these fields in the classical action of the theory.

To eliminate the self-interaction term of Eq.~(\ref{L0}), let us use the same trick applied in other four-fermion models, see e.g.~\cite{Gomes:2007mq} and references therein, that is, we introduce the  auxiliary fields $A^\mu$, $B^\mu$, and $\sigma$, so that our Lagrangian can be rewritten as
\begin{eqnarray}
\mathcal{L} &=&  \mathcal{L}_0 +\bar\psi\psi(A_\mu - g \bar{\psi} i \overset{\leftrightarrow}{\partial}_{\!\mu}\psi  / \bar{\psi} \psi)(B^\mu -  e \bar{\psi} \gamma^\mu \psi / \bar{\psi} \psi) +\sigma(\upsilon-\bar{\psi}\psi) \nonumber\\
&=&  (A_\mu B^\mu +\sigma)\upsilon + \bar\psi (i \slashed{\partial} - \sigma - e \slashed{A} - 2ig B^\mu\partial_\mu -ig(\partial_\mu B^\mu))\psi,
\end{eqnarray}
where the condition $v - \bar{\psi}\psi = 0$ is a constraint (we note that a similar constraint arises in the nonlinear sigma model, see, e.g., \cite{Girotti:2001gs} and references therein, and in the $CP^{n-1}$ models \cite{DAdda:1978dle}). It is important that this Lagrangian can be treated with the use of the standard field theory framework, being free of the controversial terms with fields in denominators. We note that, actually, for the validity of the perturbative expansion within our study, since the fields $A_{\mu}$ and $B_{\mu}$ are the external ones, it is sufficient to require these fields to be small, so, in fact, we have the expansion in $eA_{\mu}$ and $gB_{\mu}$ rather than in $e$ and $g$. 

In this section, our goal is to determine the exact potential for $A_\mu$, $B_\mu$, and $\sigma$. To this end, we begin with the generating functional
\begin{eqnarray}
Z(\bar \eta,\,\eta) &=& \int DA_\mu DB_\mu D\sigma D\psi D\bar\psi\ e^{i\int d^4x({\cal L}+\bar\eta\psi+\bar\psi\eta)}\nonumber\\
&=& \int DA_\mu DB_\mu D\sigma\ e^{i\int d^4x (A_\mu B^\mu +\sigma)\upsilon}\int D\psi D\bar\psi\ e^{i\int d^4x(\bar\psi S^{-1}\psi+\bar\eta\psi+\bar\psi\eta)},
\end{eqnarray}
where $S^{-1}=i \slashed{\partial} - \sigma - e \slashed{A} - 2ig B^\mu\partial_\mu -ig(\partial_\mu B^\mu)$ denotes the operator that defines the quadratic part of the action, that is, the inverse of the propagator. Next, we shift the fermionic fields as $\psi\rightarrow \psi-S\eta$ and $\bar{\psi}\rightarrow \bar{\psi}-\bar{\eta}S$. Under these transformations, the combination $\bar\psi S^{-1}\psi+\bar\eta\psi+\bar\psi\eta$ becomes $\bar\psi S^{-1}\psi-\bar\eta S \eta$. Consequently, we obtain
\begin{eqnarray}
Z(\bar\eta,\eta) &=& \int DA_\mu DB_\mu D\sigma\ e^{i\int d^4x (A_\mu B^\mu +\sigma)\upsilon}\int D\psi D\bar\psi\ e^{i\int d^4x(\bar\psi S^{-1}\psi-\bar\eta S \eta)}.
\end{eqnarray}
Finally, upon performing the integration over the fermionic fields, we arrive at
\begin{equation}\label{GF}
Z(\bar\eta,\eta) = \int DA_\mu DB_\mu D\sigma\ \exp\left(iS_\mathrm{eff}[A,B,\sigma] - i\int d^4x\, \bar\eta\, S\, \eta \right),
\end{equation}
where the corresponding one-loop effective action (i.e., the fermionic determinant) takes the form
\begin{equation}\label{Seff}
S_\mathrm{eff}[A,B,\sigma] = \int d^4x (A_\mu B^\mu +\sigma)\upsilon - i\mathrm{Tr}\ln(\slashed{p}-\sigma-e\slashed{A}-2gB^\mu p_\mu-ig(\partial_\mu B^\mu)).
\end{equation}
The symbol $\mathrm{Tr}$ denotes the trace over Dirac matrices, together with the integration over momentum or coordinate space. Employing the key identity of the derivative expansion method $B^\mu p_\mu = (p-i\partial)_\mu B^\mu$, which allows one to obtain the fermionic determinant as a power series in derivatives (see, e.g., \cite{Das:1987yb,Das:1987yc,Babu:1987rs}), we can rewrite the above expression as
\begin{equation}\label{Seff1}
S_\mathrm{eff}[A,B,\sigma] = \int d^4x (A_\mu B^\mu +\sigma)\upsilon - i\mathrm{Tr}\ln(\slashed{p}-\sigma-e\slashed{A}-g(2p-i\partial)_\mu B^\mu).
\end{equation}

Thus, after neglecting the terms proportional to $\partial_\mu B^\mu$, which are irrelevant for the calculation of the effective potential, and applying the shift $p_\mu \to p_\mu + e A_\mu$, which is possible as long as $A_{\mu}$ is constant, the effective potential takes the form
\begin{equation}\label{Vef}
V_\mathrm{eff} = -(A_\mu B^\mu +\sigma)\upsilon + i\mathrm{tr}\int\frac{d^4p}{(2\pi)^4}\,\ln(\slashed{p}-\sigma-2g(p+eA)_\mu B^\mu),
\end{equation}
where $\mathrm{tr}$ now denotes the trace over Dirac matrices. In what follows, we examine the possibility of deriving exact expressions for the effective potential.

For this, we rewrite the expression~(\ref{Vef}) in the form of the loop expansion:
\begin{eqnarray}\label{Vefq2}
V_{\text{eff}} &=& -M\upsilon +i\mathrm{tr}\int\frac{d^4p}{(2\pi)^4} \ln(\slashed{p}-M-2gp_\mu B^\mu) \nonumber\\
&=& -M\upsilon +i\mathrm{tr}\int\frac{d^4p}{(2\pi)^4} \ln(\slashed{p}-M) +\sum_{n=1}^{\infty}V_{\mathrm{eff}}^{(n)},
\end{eqnarray}
where $M=\sigma+A_\mu B^\mu$ and
\begin{equation}\label{Vefn}
\sum_{n=1}^{\infty}V_{\mathrm{eff}}^{(n)} = -\sum_{n=1}^{\infty}\frac{i}{n}\mathrm{tr}\int\frac{d^4p}{(2\pi)^4} \frac{(\slashed{p}+M)^n}{(p^2-M^2)^n}(2gp_\mu B^\mu)^n.
\end{equation}
Note that we have assumed the condition $2eg = 1$, which will later be used throughout the paper as the overall multiplicative factor in front of $A_\mu B^\mu$ (we remind the reader that these fields are small within our framework).

To calculate the trace in~(\ref{Vefn}), we first consider
\begin{eqnarray}
\mathrm{tr}(\slashed{p}+M)^n = \mathrm{tr}\sum_{l=0}^{n}\frac{n!}{l!(n-l)!}\slashed{p}^l M^{n-l}.
\end{eqnarray}
Observe that, once the trace is taken, only even powers of $l$ give a nonvanishing contribution. As a result, after carrying out the integration in~(\ref{Vefn}), only even powers of $n$ will survive. Therefore, upon implementing the substitutions $n \to 2n$ and $l \to 2l$ in the $D$-dimensional expression, so that $\sum\limits_{n=1}^{\infty}V_{\mathrm{eff}}^{(n)} \to \sum_{2n=2,4,\dots}^{\infty}V_{\mathrm{eff}}^{(2n)}$, we obtain
\begin{eqnarray}
\sum_{n=1}^{\infty}V_{\mathrm{eff}}^{(2n)} &=& -\sum_{n=1}^{\infty}\sum_{l=0}^{n}\frac{i}{2n}2^{D/2}\mu^{4-D}\int\frac{d^Dp}{(2\pi)^D}\frac{2n!}{2l!(2n-2l)!}\frac{p^{2l} M^{2n-2l}}{(p^2-M^2)^{2n}}(2gp_\mu B^\mu)^{2n},
\end{eqnarray}
where, to employ the dimensional regularization framework, we adopt the definition $\mathrm{tr}\,1=2^{D/2}$. 

To further simplify the integration, we can use the following replacement:
\begin{equation}
p_{\mu_1} p_{\mu_2} \cdots p_{\mu_{2n-1}} p_{\mu_{2n}} \to \frac{(D-2)!!\, p^{2n}}{(D+2n-2)!!} \sum_{\mathrm{perm}} g_{\mu_1\mu_2} \cdots g_{\mu_{2n-1}\mu_{2n}},
\end{equation}
where the sum is taken over all permutations, yielding $(2n - 1)!!$ distinct terms. Thus, after contracting these permutations with $B^{\mu_1} B^{\mu_2} \cdots B^{\mu_{2n-1}} B^{\mu_{2n}}$, we arrive at
\begin{eqnarray}
\sum_{n=1}^{\infty}V_{\mathrm{eff}}^{(2n)} &=& -\sum_{n=1}^{\infty}\sum_{l=0}^{n}\frac{i}{2n}2^{D/2}\mu^{4-D}\int\frac{d^Dp}{(2\pi)^D}\frac{2n!}{2l!(2n-2l)!}\frac{p^{2l} M^{2n-2l}}{(p^2-M^2)^{2n}} \nonumber\\
&&\times p^{2n}(4g^2B^2)^n \frac{(D-2)!!}{(D+2n-2)!!}(2n-1)!!.
\end{eqnarray}
Now, after we have finally calculated the integral and the sums, we obtain
\begin{eqnarray}\label{Vefn2}
\sum_{n=1}^{\infty}V_{\mathrm{eff}}^{(2n)} &=& \frac{2^{-\frac{D}{2}} \pi ^{1-\frac{D}{2}} \mu^{4-D} \csc\left(\frac{\pi  D}{2}\right) M^D}{D \Gamma \left(\frac{D}{2}\right)} \left[1-(1-4 g^2B^2)^{-\frac{D+1}{2}}\right].
\end{eqnarray}

Returning to Eq.~(\ref{Vefq2}) and evaluating the second term, 
\begin{eqnarray}
i\mathrm{tr}\int\frac{d^4p}{(2\pi)^4} \ln(\slashed{p}-M) &=& i2^{D/2}\mu^{4-D}\int\frac{d^Dp}{(2\pi)^D}\frac{1}{2}\ln(-p^2+M^2) \nonumber\\
&=&-\frac{2^{-\frac{D}{2}} \pi ^{1-\frac{D}{2}} \mu^{4-D} \csc\left(\frac{\pi  D}{2}\right) M^D}{D \Gamma \left(\frac{D}{2}\right)},
\end{eqnarray}
we find that it precisely cancels the first term in~(\ref{Vefn2}). Consequently, the exact expression for the effective potential takes the form 
\begin{equation}\label{Vefq2c}
V_{\text{eff}} = -M\upsilon -\frac{2^{-\frac{D}{2}} \pi ^{1-\frac{D}{2}} \mu^{4-D} \csc\left(\frac{\pi  D}{2}\right) M^D}{D \Gamma \left(\frac{D}{2}\right)} (1-4 g^2B^2)^{-\frac{D+1}{2}}.
\end{equation}

Now let us analyze the gap equations, so that, for the auxiliary fields ${\cal A}_I = \{A_\mu,B_\mu,\sigma\}$, we can verify the existence of nontrivial minima $a_I = \{a_\mu m,b_\mu,m\}$, for all these fields. Specifically, we focus on the configuration in which the auxiliary fields assume the vacuum expectation values $A_\mu = a_\mu m$, $B_\mu = b_\mu$, and $\sigma = m$, which lead to the expressions
\begin{subequations}
\begin{eqnarray}
\label{dVds}\frac{\partial V_{\mathrm{eff}}}{\partial\sigma}
\Big|_{{\cal A}_I=a_I} \hspace{-0.25cm}&=& -\upsilon-\frac{2^{-\frac{D}{2}} \pi ^{1-\frac{D}{2}} \mu^{4-D} \csc\left(\frac{\pi D}{2}\right)m^{D-1}(1+a\cdot b)^{D-1}}{\Gamma \left(\frac{D}{2}\right)}\left(1-4g^2b^2\right)^{-\frac{D+1}{2}} = 0, \hspace{1cm}\\
\label{dVdA}\frac{\partial V_{\mathrm{eff}}}{\partial A^\mu}
\Big|_{{\cal A}_I=a_I} \hspace{-0.25cm}&=& \frac{\partial V_{\mathrm{eff}}}{\partial\sigma}
\Big|_{{\cal A}_I=a_I} b_\mu = 0, \\
\label{dVdB}\frac{\partial V_{\mathrm{eff}}}{\partial B^\mu}
\Big|_{{\cal A}_I=a_I} \hspace{-0.25cm}&=& \frac{\partial V_{\mathrm{eff}}}{\partial\sigma}
\Big|_{{\cal A}_I=a_I} a_\mu m -\frac{2^{2-\frac{D}{2}} \pi ^{1-\frac{D}{2}} \mu^{4-D} \csc\left(\frac{\pi  D}{2}\right) m^D(1+a\cdot b)^D}{D \Gamma \left(\frac{D}{2}\right)} \\
&&\times (1-4 g^2b^2)^{-\frac{D+3}{2}}b_\mu = 0. \nonumber
\end{eqnarray}
\end{subequations}
Thus, by analyzing Eq.~(\ref{dVdB}), we deduce that $a \cdot b = -1$ (or, upon restoring the coupling constants, that $a \cdot b = -\frac{1}{2eg}$). From, for example, Eq.~(\ref{dVds}), we then obtain that $\upsilon = 0$. On the other hand, following~\cite{Amati:1981xt}, by imposing $A_\mu = 0$, $B_\mu = 0$, and $\sigma = m$, we obtain
\begin{equation}
\upsilon = -\frac{2^{-\frac{D}{2}} \pi^{1-\frac{D}{2}} \mu^{4-D} \csc\left(\frac{\pi D}{2}\right) m^{D-1}}{\Gamma\left(\frac{D}{2}\right)},
\end{equation}
which allows us to conclude that the fermion mass is dynamically generated.

To finalize our analysis, we expand the expression~(\ref{Vefq2c}) around $D=4$, with $\sigma=m$ and $\upsilon=0$, and perform the rescaling $A_\mu \to A_\mu m$, which yields
\begin{equation}
V_{\text{eff}} = \frac{(1+2egA\cdot B)^4m^4}{8\pi^2(1-4g^2B^2)^{5/2}}\left(\frac{1}{\epsilon}-\ln\frac{m}{\mu'}-\frac{1}{2}\ln\frac{(1+2egA\cdot B)^2}{(1-4g^2B^2)}+\frac{3}{4}\right),
\end{equation}
where $\epsilon = 4 - D$ and $\mu'^2 = 4\pi\mu^2 e^{-\gamma}$. Next, we define the renormalized mass $m_R = Z_m^{-1/4} m$, with
\begin{equation}
\frac{1}{Z_{m}} = \frac{1}{8\pi^2\epsilon'},
\end{equation}
and $\frac{1}{\epsilon'} = \frac{1}{\epsilon} - \ln\frac{m}{\mu'}$. With these definitions, the effective potential becomes
\begin{equation}
V_{\text{eff}} = \frac{(1+2egA\cdot B)^4m_R^4}{(1-4g^2B^2)^{5/2}}.
\end{equation}
Analyzing the above expression, we observe that, with $A_\mu=a_\mu$, the potential is non-negative and reaches a nontrivial global minimum when $a\cdot B=-\frac{1}{2eg}$, subject to the reality condition $1-4g^2B^2>0$. For a spacelike $B_\mu$, this reality condition is automatically satisfied, and the minimum $V_{\mathrm{eff}}=0$ is therefore manifest. By contrast, for a timelike $B_\mu$, the existence of such a minimum is conditional: in addition to $a\cdot B=-\frac{1}{2eg}$, one must also require $B^2<\frac{1}{4g^2}$ to ensure that the effective potential remains real and finite. This result agrees with Ref.~\cite{Garavaglia:1986cy}.

For the contribution up to fourth order in $B_\mu$, we obtain
\begin{equation}
V_{\mathrm{eff}}=m_R^4\bigl(8eg\,A\cdot B+10g^2B^2+24e^2g^2(A\cdot B)^2+80eg^3(A\cdot B)B^2+70g^4B^4\bigr),
\end{equation}
where the constant term $m_R^4$ has been omitted. Setting $A_\mu=a_\mu$ and analyzing timelike and spacelike $B_\mu$ separately, we find that the quartic potential is bounded from below and exhibits a nontrivial global minimum in both cases, without requiring any additional conditions on $B^2$.

\section{Effective action}\label{action}

In this section, we study the one-loop low-energy effective action for the vector fields $A_\mu$ and $B_\mu$ by setting $\sigma = m$ and $\upsilon = 0$, and rescaling $A_\mu \to m A_\mu$, that is, besides the effective potential, we also obtain kinetic terms for both of our vector fields. To this end, we rewrite Eq.~(\ref{Seff1}) as follows:
\begin{eqnarray}
S_{\mathrm{eff}}[A,B]
&=& -i\mathrm{Tr}\ln(\slashed{p}-m) -i\mathrm{Tr}\ln[1-S(p)\Gamma^I(p,i\partial) {\cal A}_I] \nonumber\\
&=& -i\mathrm{Tr}\ln(\slashed{p}-m) +\sum_{n=1}^{\infty}S_{\mathrm{eff}}^{(n)}[A,B], 
\end{eqnarray}
where
\begin{equation}
S_{\mathrm{eff}}^{(n)}[A,B]=\frac{i}{n}\mathrm{Tr}[S(p)\Gamma^I(p,i\partial) {\cal A}_I]^n    
\end{equation}
and now $S^{-1}(p)=\slashed{p}-m$ is the inverse propagator and ${\cal A}_I=\{A_\mu,B_\mu\}$ is the set of dynamical fields, with $\Gamma^I(p,i\partial)=\{em\gamma^\mu,g(2p-i\partial)^\mu\}$. 

Within our calculations, let us proceed in a way similar to \cite{Gomes:2007mq}. First of all, we note that the $n = 1$ (tadpole) and $n = 3$ (three-point) contributions evidently vanish, i.e., $S_{\mathrm{eff}}^{(1)}[A,B]=0$ and $S_{\mathrm{eff}}^{(3)}[A,B]=0$. We therefore focus on the contributions with $n = 2$ and $n = 4$, whose analysis is sufficient to generate the kinetic and lower-order potential terms. First, for $n=2$, we have
\begin{eqnarray}
S^{(2)}_\mathrm{eff} &=& \frac{i}{2}\int d^4x\,\Pi^{IJ}\mathcal{A}_I\mathcal{A}_J,
\end{eqnarray}
where
\begin{equation}
\Pi^{IJ}=\mathrm{tr}\int\frac{d^4p}{(2\pi)^4}\,S(p)\Gamma^I(p,k)S(p-k)\Gamma^J(p,k),
\end{equation}
with, for simplicity, $k_\mu=i\partial_\mu$. Note that we have assumed $\Gamma^J(p-k,k')=\Gamma^J(p,k)$, which follows from momentum conservation, $k_\nu + k'_\nu = 0$. Here, $i\partial_\mu$ acts on $\mathcal{A}_I=\{A_\mu,B_\mu\}$, and $i\partial'_\nu$ acts on $\mathcal{A}_J=\{A_\nu,B_\nu\}$.

After applying Feynman parametrization, evaluating the trace and the integral of the above expression, and using $S^{(2)}_{\mathrm{eff}}[A,B]=\int d^4x\, {\cal L}^{(2)}_{\mathrm{eff}}$, we obtain
\begin{equation}
{\cal L}^{(2)}_{\mathrm{eff}}={\cal L}(A)+{\cal L}(A,B)+{\cal L}(B),
\end{equation}
where
\begin{equation}
{\cal L}(A) = \left[-\frac{e^2m^2}{12\pi^2\epsilon'}+\frac{e^2m^2}{72\pi^2k^4(4m^2k^2)}{\cal F}(k,m)\right]\left(k^2 g^{\mu\nu}-k^\mu k^\nu\right)A_\mu A_\nu,
\end{equation}
\begin{eqnarray}
{\cal L}(A,B) &=& 
\left[
\frac{egm^2}{12\pi^2\epsilon'}
\left(k^2 g^{\mu\nu}-k^\mu k^\nu\right)-\frac{egm^4}{2\pi^2\epsilon'}\,g^{\mu\nu}\right. \nonumber\\
&&\left.+\frac{egm^2}{36\pi^2 k^2}
(g^{\mu\nu}\,\mathcal{G}(k,m)+k^\mu k^\nu\,\mathcal{H}(k,m))
\right](A_\mu B_\nu + B_\mu A_\nu),
\end{eqnarray}
\begin{eqnarray}
{\cal L}(B) &=& 
\bigg\{-\frac{1}{12\pi^2 \epsilon'}
\left(
\frac{g^2k^2}{2}
+5g^2m^2
\right)
\left(k^2 g^{\mu\nu}-k^\mu k^\nu\right)
-
\frac{5g^2m^4}{4\pi^2\epsilon'}\,g^{\mu\nu}
+
\frac{g^2}{144\pi^2 k^2}\nonumber\\
&&\times
\bigg[
\frac{6(4m^2-k^2)^{5/2}}{\sqrt{k^2}}
\csc^{-1}\left(\frac{2m}{\sqrt{k^2}}\right)
\left(k^2 g^{\mu\nu}-k^\mu k^\nu\right)+
(-8k^6+74k^4m^2 \nonumber\\
&&-195k^2m^4)g^{\mu\nu}+
2\left(
4k^4-37k^2m^2+48m^4
\right)
k^\mu k^\nu
\bigg]\bigg\}B_\mu B_\nu,
\end{eqnarray}
with
\begin{eqnarray}    
{\cal F}(k,m)&=&5k^6-8k^4m^2-48k^2m^4-6\sqrt{k^2(4m^2-k^2)} \nonumber\\
&&\times\left[k^4-8m^4-2m^2k^2\csc^{-1}\!\left(\frac{2m}{\sqrt{k^2}}\right)
\right], \\
{\cal G}(k,m)&=&4k^4-21k^2m^2+3\sqrt{k^2}\,(4m^2-k^2)^{3/2}\csc^{-1}\!\left(\frac{2m}{\sqrt{k^2}}\right), \\
{\cal H}(k,m)&=&4(3m^2-k^2)-3\frac{(4m^2-k^2)^{3/2}}{\sqrt{k^2}}\csc^{-1}\!\left(\frac{2m}{\sqrt{k^2}}\right).
\end{eqnarray}
Thus, in the low-energy limit ($k^2\ll m^2$), we have
\begin{eqnarray}
\mathcal{L}_{\text{eff}}^{(2)}&=& 
\frac{e^2 m^2}{12 \pi^2 \epsilon'}  A_\mu(\partial^2 g^{\mu \nu} - \partial^{\mu}\partial^\nu)A_\nu 
-\frac{egm^2}{6 \pi^2\epsilon'} A_\mu(\partial^2 g^{\mu \nu} - \partial^\mu \partial^\nu )B_\nu \nonumber\\
&&-\frac{5g^2m^2}{12 \pi^2}  \left ( \frac{1}{\epsilon'} + \frac{3}{10} \right )  B_\mu(\partial^2 g^{\mu \nu} - \partial^\mu \partial^\nu)B_\nu -\frac{eg m^4}{\pi^2}  \left ( \frac{1}{\epsilon'} + \frac{1}{2} \right ) A_\mu B^\mu  \nonumber\\
&&-\frac{5 g^2 m^4}{4 \pi^2}  \left (\frac{1}{\epsilon'} + \frac{11}{20}  \right ) B_\mu B^\mu +{\cal O}(\partial^4). 
\end{eqnarray}
Then, by considering the renormalization constants
\begin{equation}
\frac{1}{Z_A}=\frac{e^2 m^2}{6 \pi^2 \epsilon'}\quad \text{and}\quad \frac{1}{Z_B}=\frac{5g^2 m^2}{6 \pi^2 \epsilon'},
\end{equation}
we can define the renormalized fields $A_R^\mu=Z_A^{-1/2}A_\mu$ and $B_R^\mu=Z_B^{-1/2}B_\mu$, as well as the coupling constants $e_R=Z_A^{1/2}e$ and $g_R=Z_B^{1/2}g$. We can then rewrite the above Lagrangian as
\begin{eqnarray}
\mathcal{L}_{\text{eff}}^{(2)} &=& -\frac{1}{4}F_{R\mu\nu}(A)F_R^{\mu\nu}(A)+\frac{1}{4}F_{R\mu\nu}(B)F_R^{\mu\nu}(B)+\frac{1}{\sqrt{5}}F_{R\mu\nu}(A)F_R^{\mu\nu}(B) \nonumber\\
&&- 8e_R g_R m_R^4 A_{R\mu} B_R^{\mu} -10 g_R^2 m_R^4 B_{R\mu} B_R^{\mu},
\end{eqnarray}
where $F_{\mu\nu}(A)=\partial_\mu A_\nu-\partial_\nu A_\mu$ and $F_{\mu\nu}(B)=\partial_\mu B_\nu-\partial_\nu B_\mu$, and we have omitted higher-order terms.

Now, for $n=4$, we have
\begin{eqnarray}
S^{(4)}_\mathrm{eff} &=& \frac{i}{4}\int d^4x\,\Pi^{IJKL}\mathcal{A}_I\mathcal{A}_J\mathcal{A}_K\mathcal{A}_L,
\end{eqnarray}
where
\begin{equation}
\Pi^{IJKL}=\mathrm{tr}\int\frac{d^4p}{(2\pi)^4}\,S(p)\Gamma^I(p,0)S(p)\Gamma^J(p,0)S(p)\Gamma^K(p,0)S(p)\Gamma^L(p,0) +{\cal O}(\partial^4).
\end{equation}
Then, after performing the trace and integral (and omitting the higher-order terms), the resulting fourth-order eﬀective Lagrangian is
\begin{eqnarray}
\mathcal{L}_{\text{eff}}^{(4)} &=& - \frac{3e^2 g^2m^4}{\pi^2} \left( \frac{1}{\epsilon'} + \frac{1}{6} \right) A_\mu B^\mu A_\nu B^\nu -\frac{10e g^3m^4}{\pi^2} \left(\frac{1}{\epsilon'}  +\frac{3}{10}  \right) A_\mu B^\mu B_\nu B^\nu \nonumber \\
&& - \frac{35 g^4 m^4}{4\pi^2} \left( \frac{1}{\epsilon'} + \frac{57}{140} \right) B_\mu B^\mu B_\nu B^\nu.
\end{eqnarray}
Therefore, by rewriting the above Lagrangian in terms of the normalized quantities, we obtain
\begin{eqnarray}
{\cal L}_{\mathrm{eff}}^{(4)} &=& -24e_R^2g_R^2m_R^4(A_{R\mu} B_R^\mu)^2 -80e_Rg_R^3m_R^4(A_{R\mu} B_R^{\mu})(B_{R\nu} B_R^\nu) -70g_R^4m_R^4(B_{R\mu  }B_R^\mu)^2. 
\end{eqnarray}

Finally, collecting the quadratic and quartic contributions, the renormalized low-energy effective Lagrangian associated with the vector fields $A_\mu$ and $B_\mu$ takes the form
\begin{eqnarray}
{\cal L}_\mathrm{eff} &=& -\frac{1}{4}F_{R\mu\nu}(A)F_R^{\mu\nu}(A)+\frac{1}{4}F_{R\mu\nu}(B)F_R^{\mu\nu}(B)+\frac{1}{\sqrt{5}}F_{R\mu\nu}(A)F_R^{\mu\nu}(B) \nonumber\\
&& -8e_R g_R m_R^4 A_{R\mu} B_R^{\mu} -10 g_R^2 m_R^4 B_{R\mu} B_R^{\mu} -24e_R^2g_R^2m_R^4(A_{R\mu} B_R^\mu)^2 \nonumber\\
&&-80e_Rg_R^3m_R^4(A_{R\mu} B_R^{\mu})(B_{R\nu} B_R^\nu) -70g_R^4m_R^4(B_{R\mu  }B_R^\mu)^2. 
\end{eqnarray}
As a result, we succeeded in generating the low-energy effective action, involving kinetic terms and a potential up to the fourth order in the fields. As we argued above, the potential is bounded from below and possesses a nontrivial minimum.

\section{Summary}\label{summary}

We proposed a new model allowing for a dynamical Lorentz symmetry breaking. Unlike earlier theories, our model allows one to generate kinetic terms for two distinct vector fields with one of them displaying a gauge symmetry. Our starting point is the highly non-polynomial spinor field theory proposed in \cite{Garavaglia:1986cy}. The vector fields are introduced as Lagrange multipliers allowing one to rewrite the action in a conventional form without fields in denominators. Afterwards, the spinors are integrated out, and the one-loop low-energy effective action (LEEA), represented by the fermionic determinant, is calculated explicitly. The most interesting term of this LEEA is the effective potential, which is bounded from below. If one of the vector fields is fixed to be equal to its v.e.v., the effective potential for the other field allows for a nontrivial minimum, implying therefore the possibility of a dynamical LSB. Besides this, the Maxwell kinetic terms are generated for both of our vector fields. Effectively, our study represents an example of generating the bumblebee model as a particular case of a more generic theory which, in principle, could serve as a possible explanation of the origin of the bumblebee theory.

A possible extension of this model could consist in its coupling to other fields, for example, gauge and scalar ones, which could allow a dynamical generation of couplings of these fields with the bumblebee one, providing thus a consistent mechanism for the emergence of LV extensions for theories involving these fields. Other possible extensions could consist in the introduction of $N$ different spinor fields, providing hence a $U(N)$ symmetry, with further consideration of the effective dynamics of our condensate vector fields within the framework of the $1/N$ expansion, as well as in the introduction of a finite temperature, which is especially interesting taking into account that our vector fields actually are fermion condensates. Thus, we can study possible phase transitions. We expect to perform these studies in forthcoming papers.

{\bf Acknowledgments.}  T.~M. would like to thank CNPq (Project No. 309360/2025-0) and FAPEAL (Project No. E:60030.0000002341/2022) for financial support. The work of A.~Yu.~P. has been partially supported by CNPq (project No. 303777/2023-0).

\end{document}